\documentclass{PoS}
\usepackage{graphicx}
\usepackage{wrapfig}
\usepackage{wrapfig,booktabs}
\usepackage{color}
\usepackage{multirow}
\usepackage{array}
\usepackage{float}
\usepackage{amsmath,amssymb,amsfonts,amstext,amsthm}
\newcommand{\bea}{\begin{eqnarray}}
\newcommand{\eea}{\end{eqnarray}}
\newcommand{\BAN}{\begin{eqnarray*}}
\newcommand{\EAN}{\end{eqnarray*}}

\newcommand{\p}{\partial}
\newcommand{\pslash}{p\kern-1ex /}
\newcommand{\lslash}{l\kern-1ex /}
\newcommand{\kslash}{k\kern-1ex /}
\newcommand{\dslash}{\p\kern-1.2ex /}
\newcommand{\Dslash}{{\cal D}\kern-1.5ex /}

\newcommand{\tr}{{\rm tr}}

\newcommand{\Dodwf}{\mathcal{D}}
\def\u{{\bf u}}
\def\d{{\bf d}}
\def\s{{\bf s}}
\def\c{{\bf c}}
\def\b{{\bf b}}
\def\t{{\bf t}}
\def\q{{\bf q}}

\def\ubar{\bar{\bf u}}
\def\dbar{\bar{\bf d}}

\def\qbar{\bar{\bf q}}

\title{Simulation of dynamical $(\u,\d,\s,\c)$ domain-wall/overlap quarks at the physical point }

\ShortTitle{Lattice QCD with physcial (u,d,s,c) domain-wall quarks}

\author{\speaker{Ting-Wai Chiu}\thanks{This work is supported by the Ministry of Science and Technology
(Nos.~107-2119-M-003-008, 105-2112-M-002-016, 102-2112-M-002-019-MY3), 
and National Center for Theoretical Sciences (Physics Division).} \ (TWQCD Collaboration) \\
        Institute of Physics, Academia Sinica, Taipei, Taiwan 11529, R.O.C.\\
        Physics Department, National Taiwan Normal University, Taipei, Taiwan 11677, R.O.C.\\
        Physics Department, National Taiwan University, Taipei, Taiwan 10617, R.O.C.\\
        E-mail: \email{twchiu@phys.ntu.edu.tw}
}

\abstract{We perform hybrid Monte-Carlo simulation of $N_f=2+1+1 $ lattice QCD with 
domain-wall quarks at the physical point. 
The simulation is carried out on the $ L^3 \times T = 64^3 \times 64 $ lattice with lattice spacing 
$a \sim 0.064 $~fm ($ L > 4 $ fm, and $ M_\pi L > 3 $), 
using the Nvidia DGX-1 (8 V100 GPUs interconnected by the NVLink). 
To attain the maximal chiral symmetry for a finite extent ($N_s=16$) in the fifth dimension, 
we use the optimal domain-wall fermion for the quark action, 
together with the exact one-flavor action for domain-wall fermion. 
We outline the salient features of our simulation (without topology freezing, and small residual mass), 
together with the preliminary result of the masses of $ \pi^\pm $, $ K^\pm $, and $ D^\pm $.
}

\FullConference{The 36th Annual International Symposium on Lattice Field Theory - LATTICE2018\\
		22-28 July, 2018\\
		Michigan State University, East Lansing, Michigan, USA.}

\begin{document}

\section{Introduction}

The holy grail of lattice QCD is to simulate QCD with all quarks at their physical masses, with 
sufficiently large volume and fine lattice spacing, then to extract physics from these gauge ensembles. 
Since the $\t$ quark is extremely short-lived and it decays to W-boson and
$ \b/\s/\d $ quarks before it can interact with other quarks through the gluons, 
it can be neglected in QCD simulations.
Even after neglecting the $\t$ quark, to simulate $(\u, \d, \s, \c, \b)$ quarks
at their physical masses (ranging from $\sim 3-4500 $~MeV) is still a very challenging problem, 
For example, if one designs the simulation on a $ L^4 $ lattice with $ M_\pi L > 4 $ 
and lattice spacing $ a < 0.035 $~fm such that the $\b$ quark mass 
satisfies $ m_b a < 0.8 $, then it would require $ L/a > 180 $, 
which is beyond the capability of the current generation of supercomputers with $\sim 100-200 $~Petaflops/s. 

\begin{wrapfigure}{r}{0.4\textwidth}
\centering
\includegraphics[width=6cm,clip=true]{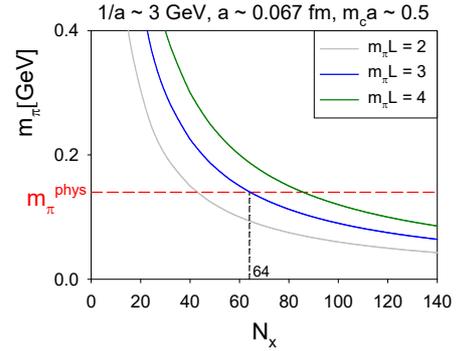}
\caption{\label{fig:design_LQCD_udsc}Design lattice QCD with physical $(\u,\d,\s,\c)$ quarks.
}
\end{wrapfigure}

Now if we neglect the $ \b $ quark, then lattice QCD with physical $ (\u,\d,\s,\c) $ quarks can be 
simulated on a $ 64^4 $ lattice with lattice spacing $ a \sim 0.067 $~fm, 
which satisfies $ m_c a < 0.6 $, $ M_\pi L > 3 $, and $ L > 4$~fm, 
as shown in Fig. \ref{fig:design_LQCD_udsc}.      
The hybrid Monte-Carlo (HMC) \cite{Duane:1987de} simulation of this $ N_f = 2 + 1 + 1 $ lattice QCD 
can be performed with the current generation of supercomputers,  
however, the degree of difficulties depends on which kind of lattice fermion is used.

It took 24 years (1974-1998) to realize that lattice QCD with exact chiral symmetry 
\cite{Kaplan:1992bt,Neuberger:1997fp}
is the ideal theoretical framework to study the nonperturbative physics from the first principles of QCD.
Howeve, it is challenging to perform the Monte Carlo simulation such that the chiral symmetry 
is preserved to a very high precision and all topological sectors are sampled ergodically.
Moreover, the computational requirement for lattice QCD with domain-wall quarks on a 5-dimensional lattice 
is 10-100 times more than their counterparts with traditional lattice fermions (e.g., 
Wilson, staggered, and their variants). 

In this talk, I report the first HMC simulation of lattice QCD with 
physical $N_f=2+1+1 $ domain-wall quarks. The simulation is performed 
on a $ 64^4 $ lattice with the extent $ N_s = 16 $ in the fifth dimension, 
using just one GPU server on the table top, 
the Nvidia DGX-1 (with eight V100 GPUs interconnected by the NVLink,  
and total device memory $16 \times 8 = 128$~GB).
In general, to simulate $N_f = 2+1+1$ lattice QCD a $ 64^4 $ lattice 
with any lattice Dirac operator $ D $ usually requires memory much larger than 128~GB, 
since each one-flavor pseudofermion action is expressed as the rational approximation 
of $ \Phi^\dagger (D^\dagger D)^{-1/2} \Phi $, 
requiring a large number (proportional to the number of poles in the rational approximation) 
of long vectors in computing the fermion forces in the molecular dynamics. 
However, for domain-wall fermion, one can use the exact one-flavor pseudofermion action (EOFA) 
with a positive-definite and Hermitian Dirac operator \cite{Chen:2014hyy}.
Moreover, using EOFA for $ N_f = 2 + 1 +1 $ lattice QCD with domain-wall quarks 
not only saves the memory such that the HMC simulation on a $64^4 \times 16 $ lattice 
can be fitted into 128 GB device memory of Nvidia DGX-1, 
but also enhances the HMC efficiency significantly. 

\section{Generation of the gauge configurations}

As pointed out in Ref. \cite{Chen:2017kxr}, for domain-wall fermions, 
to simulate $ N_f = 2 +1 + 1 $ 
amounts to simulate $ N_f = 2 + 2 + 1 $, according to the identity
\bea
\label{eq:Nf2p1p1}
&& \frac{\det \Dodwf(m_{u/d})}{\det \Dodwf(m_{PV})} 
   \frac{\det \Dodwf(m_{u/d})}{\det \Dodwf(m_{PV})} 
   \frac{\det \Dodwf(m_s)}{\det \Dodwf(m_{PV})}
   \frac{\det \Dodwf(m_c)}{\det \Dodwf(m_{PV})}    \\
&=&
\label{eq:Nf2p2p1_B}
\left( \frac{\det \Dodwf(m_{u/d})}{\det \Dodwf(m_{PV})} \right)^2
\left( \frac{\det \Dodwf(m_c)}{\det \Dodwf(m_{PV})} \right)^2
\frac{\det \Dodwf(m_s)}{\det \Dodwf(m_{c})},
\eea
where $ \Dodwf(m_q) $ denotes the domain-wall fermion operator with bare quark mass $ m_q $,
and $ m_{PV} $ the mass of the Pauli-Villars field. Since the simulation of 2-flavors
is faster than the simulation of one-flavor, it is better to simulate $ N_f = 2 + 2 + 1 $
than $ N_f = 2 + 1 + 1 $. In this study, we use (\ref{eq:Nf2p2p1_B}) for our HMC simulations.

For the gluon fields, we use the Wilson plaquette gauge action at $ \beta = 6/g_0^2 = 6.20 $.
For the quark fields, we use the optimal domain-wall fermion \cite{Chiu:2002ir}, where 
the optimal weights $\{\omega_s, s = 1, \cdots, N_s \} $ has the $ R_5 $ symmetry \cite{Chiu:2015sea}. 
Then the effective 4-dimensional lattice Dirac operator is exactly
equal to the ``shifted" Zolotarev optimal rational approximation
of the overlap operator, with the approximate sign function $ S(H) $
satisfying the bound $ 0 \le 1-S(\lambda) \le 2 d_Z $
for $ \lambda^2 \in [\lambda_{min}^2, \lambda_{max}^2] $,
where $ d_Z $ is the maximum deviation $ | 1- \sqrt{x} R_Z(x) |_{\rm max} $ of the
Zolotarev optimal rational polynomial $ R_Z(x) $ of $ 1/\sqrt{x} $
for $ x \in [1, \lambda_{max}^2/\lambda_{min}^2] $, with degrees $ (n-1,n) $ for $ N_s = 2n $.

For the two-flavor parts, we use the pseudofermion action for 2-flavors of 
optimal domain-wall quarks, as defined in Eq. (14) of Ref. \cite{Chiu:2013aaa}.
For the one-flavor part, we use the exact pseudofermion action for one-flavor domain-wall fermion,
as defined by Eq. (23) of Ref. \cite{Chen:2014hyy}.  
The parameters of the pseudofermion actions are fixed as follows.
For the $\Dodwf(m_q) $ defined in Eq. (2) of Ref. \cite{Chiu:2013aaa},
we fix $ c = 1, d = 0 $, $ m_0 = 1.3 $, $ N_s = 16 $, and $ \lambda_{max}/\lambda_{min} = 6.20/0.05 $.
Thus $ H = H_w $ and $ m_{PV} = 2.6 $. 

We perform the HMC simulation of (2+1+1)-flavors QCD on the $ L^3 \times T = 64^3 \times 64$ lattice,
with the quark masses $ m_{\u/\d} a = 0.00125$,
$ m_{\s} a = 0.04 $, and $ m_{\c} a = 0.55 $,
which are fixed by the masses of the charged pion 
$ \pi^{\pm}(140) $, and the vector mesons $ \phi(1020) $ and $ J/\psi(3097) $ respectively.
The algorithm for simulating 2-flavors of optimal domain-wall quarks 
has been outlined in Ref. \cite{Chiu:2013aaa},
while the exact one-flavor algorithm (EOFA) for domain-wall fermions 
has been presented in Ref. \cite{Chen:2014hyy}.
In the molecular dynamics, we use the Omelyan integrator \cite{Omelyan:2001},
and the Sexton-Weingarten multiple-time scale method \cite{Sexton:1992nu}.
Moreover, we introduce auxiliary heavy fermion fields with masses $ m_H $ ($ m_q \ll m_H \ll m_{PV} $)
similar to the case of the Wilson fermion \cite{Hasenbusch:2001ne}, the so-called mass preconditioning.
For the 2-flavors parts, mass preconditioning is only applied to
the $\u/\d$ quark factor of (\ref{eq:Nf2p2p1_B}) with two levels of heavy masses 
$ m_{H_1} a = 0.01 $ and $ m_{H_2} a = 0.1 $. 
For the one-flavor part, a novel mass preconditioning has been devised for the EOFA \cite{Chen:2017gsk}
which is $\sim 20\% $ faster than the mass preconditioning we have used in Ref. \cite{Chen:2014hyy}. 
Also, based on the fact that in EOFA the fermion force of the $ \phi_1 $ field is much smaller than 
that of the $ \phi_2 $ field,
the gauge momentum updating by these two forces can be set at two different time scales.

\begin{wrapfigure}{r}{0.5\textwidth}
\centering
\includegraphics[width=8cm,clip=true]{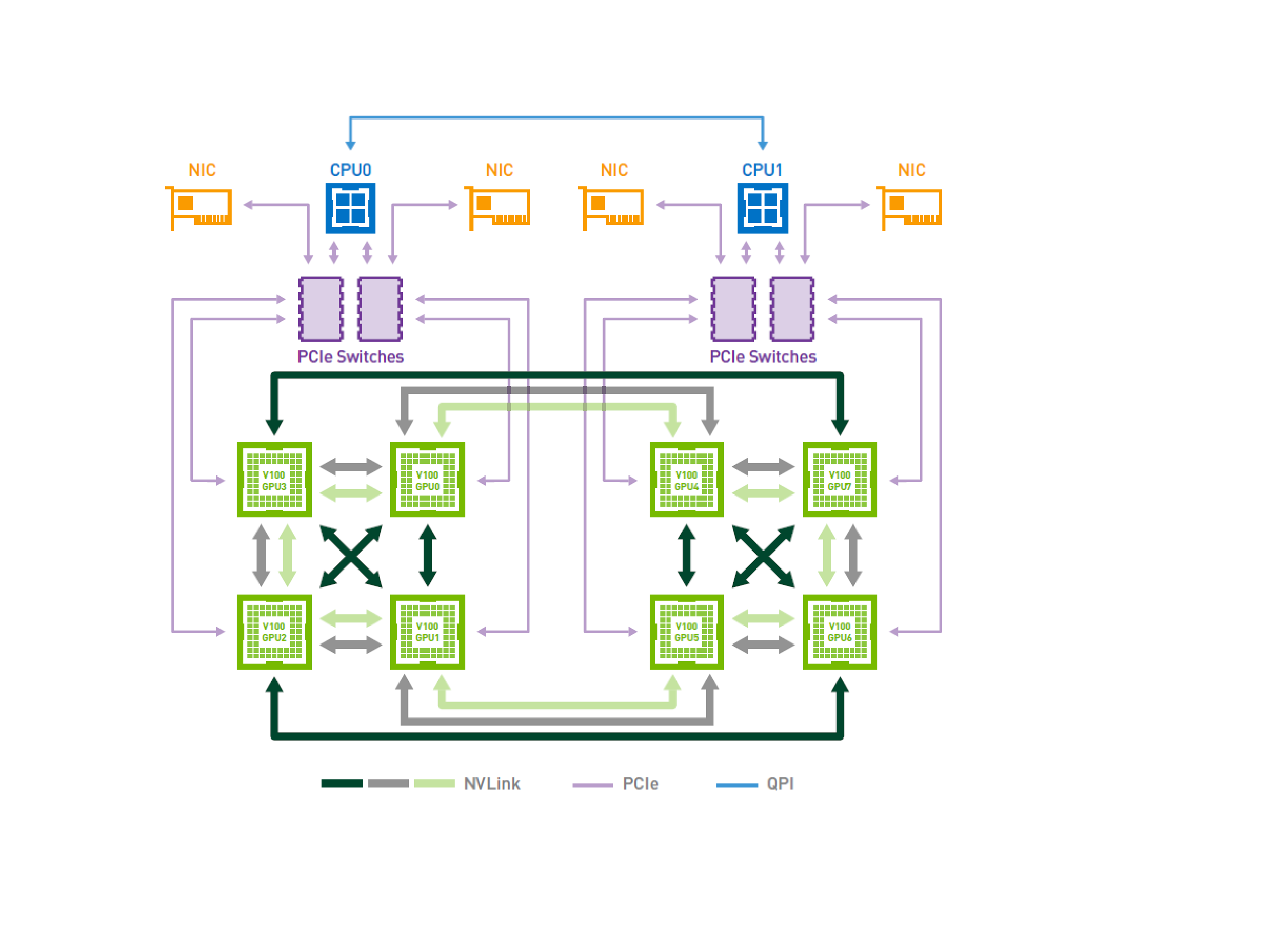}
\caption{\label{fig:DGX-1_diagram}Schematic diagram of DGX-1 from the White Paper
 ``NVIDIA DGX-1 With Tesla V100 System Architecture" (Nvidia Corporation).
}
\end{wrapfigure}

The simulation is performed with one GPU server, Nvidia DGX-1, 
with eight V100 GPUs interconnected by the NVLink, and total device memory $128$~GB. 
The schematic diagram of DGX-1 is shown in Fig. \ref{fig:DGX-1_diagram}.  
Each NVLink provides a point-to-point connections with 
data rate of 25 Gbit/s per data lane per direction, and the total data rate for the Nvidia DGX-1 
is 300 GByte/s for the total system, sum of input and output data streams.
The NVLink plays the crucial role in reducing the bottleneck in data communication between GPUs
such that the performance of the entire system can attain almost the perfect scaling with the 
number of GPUs. For conjugate gradient with 8 GPUs under OpenMP, 
we export OpenMP environment to re-map the 8 GPUs in a circle 
such that each GPU can access its neighbors P2P (peer-to-peer) through NVLinks, 
and also each CPU handles 4 GPUs. 
Our HMC simulation code attains 10 Tflops/s (sustained) on Nvidia DGX-1, and each HMC trajectory 
takes about one day. 
We have generated about 50 HMC trajectories after thermalization.  
Then sampling one configuration every 5 trajectories, we have 10 configurations for the 
preliminary measurement of physical observables.   

To determine the lattice scale, we use the Wilson flow \cite{Narayanan:2006rf,Luscher:2010iy} 
with the condition
\BAN
\label{eq:t0}
\left. \{ t^2 \langle E(t) \rangle \} \right|_{t=t_0} = 0.3,
\EAN
and obtain $ \sqrt{t_0}/a = 2.2285(18) $ for 50 HMC trajectories. 
Using $ \sqrt{t_0} =  0.1416(8) $~fm obtained by 
the MILC Collaboration for the $(2+1+1)$-flavors QCD \cite{Bazavov:2015yea},
we have $ a^{-1} = 3.104 \pm 0.018 $~GeV. 

\begin{wraptable}{r}{0.55\textwidth}
\caption{The residual masses of $ \u/\d $, $\s $, and $ \c $ quarks.}
\begin{tabular}{cccc} \\\toprule
quark & $m_q a $ & $ m_{res} a $ & $ m_{res}$~[MeV]   \\\midrule
$\u/\d$ & 0.00125  & $ 3.61(22) \times 10^{-5} $ & 0.11(1)   \\ 
$\s$    & 0.040    & $ 1.09(14) \times 10^{-5} $ & 0.03(4)   \\
$\c$    & 0.550    & $ 0.42(14) \times 10^{-5} $ & 0.01(4)   \\  \bottomrule
\end{tabular}
\label{tab:mres}
\end{wraptable} 

\section{Characteristics of the simulation}

First, we investigate the effects of chiral symmetry breaking due to finite $N_s=16$, 
by computing the residual mass of each quark flavor, 
according to the formula derived in Ref. \cite{Chen:2012jya}. 
The residual masses
of $ \u/\d$, $\s $, and $\c $ quarks are listed in Table \ref{tab:mres}.
We see that the residual mass of the $ \u/\d $ quark is $\sim 3$\% of its bare mass, amounting to
$0.11(1) $~MeV, which is expected to be smaller than other systematic uncertainties.
The residual masses of $ \s $ and $ \c $ quarks are even smaller, $ 0.03(4)$~MeV, and $0.01(4) $~MeV
respectively. This demonstrates that the optimal DWF can preserve the chiral symmetry 
to a high precision, for both light and heavy quarks.   

Next, we examine the evolution of the topological charge ($ Q_{\text{top}} $) in our HMC simulation. 
Using the Atiyah-Singer index theorem, the topological charge can be measured by the 
index of the massless overlap-Dirac operator. However, to project the zero modes 
of the massless overlap-Dirac operator for the $ 64^4 $ lattice is prohibitively expensive.
On the other hand, the clover topological charge  
$ Q_{\text{clover}} = \sum_x \epsilon_{\mu\nu\lambda\sigma} 
\tr[ F_{\mu\nu}(x) F_{\lambda\sigma}(x) ]/(32 \pi^2) $
is not reliable [where the matrix-valued field tensor $ F_{\mu\nu}(x) $ is obtained from 
the four plaquettes surrounding $ x $ on the ($\hat\mu,\hat\nu$) plane],    
unless the gauge configuration is sufficiently smooth. The smoothness can be attained 
by the Wilson flow which is a continuous-smearing process to average gauge field 
over a spherical region of root-mean-square radius $ R_{rms} = \sqrt{8 t} $, where $ t $ is 
the flow time. Theoretically, if a gauge configuration 
has been flowed for a sufficiently long time to satisfy 
the condition \cite{Luscher:2010iy,Luscher:1981zq,Phillips:1986qd}
\bea 
\label{eq:admissibility}
\min_{\text{plaq}}\left\{\frac{1}{3}{\text{Re}} \ \tr \ U_{\text{plaq}}\right\} >\frac{44}{45} \simeq 0.978, 
\eea 
\begin{wrapfigure}{r}{0.55\textwidth}
\centering
\includegraphics[width=8cm,clip=true]{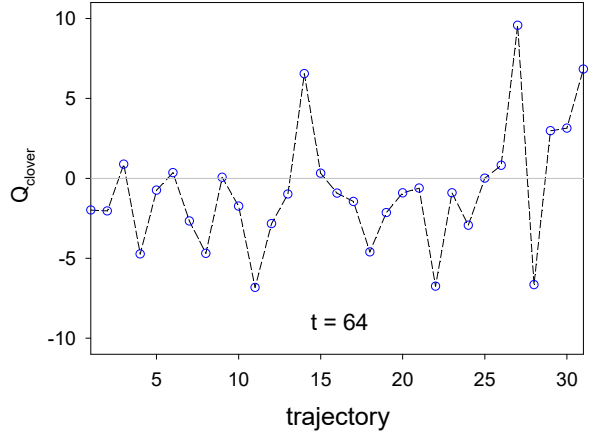}
\caption{\label{fig:Q_traj} The evolution of the topological charge of 31 successive HMC trajectories.
}
\end{wrapfigure}
then $ Q_{\text{clover}} $ would be close to an integer, and 
further Wilson flow acting on this configuration would not change the integer part of $ Q_{\text{clover}} $.
In other words, applying the Wilson flow to an ensemble of lattice gauge configurations
for a sufficiently long time 
can let them fall into topological sectors, similar to the gauge fields in the continuum theory.
In practice, some configuration may take a very long flow time (e.g., $ t > 128 $) 
in order to satisfy the condition (\ref{eq:admissibility}), especially for a large lattice like $ 64^4 $.     
We observe that if the flow equation is integrated from $ t = 0 $ to $ t = 64 $ with $ \Delta t = 0.01 $, 
then the condition (\ref{eq:admissibility}) is satisfied by all HMC trajectories generated so far, 
and the integer value of $ Q_{\text{clover}} $ becomes a constant for a long interval of flow time 
before $ t=64 $, thus $ Q_{\text{top}} = [Q_{\text{clover}}] $ is well-defined. 
In Fig.~\ref{fig:Q_traj}, the evolution of $ Q_{\text{clover}}(t=64)$ of 31 successive HMC trajectories 
is plotted.  Evidently, the HMC simulation does not suffer from topology freezing, and will likely 
sample all topological sectors ergodically when the number of trajectories becomes sufficiently large.

\section{Preliminary result of the masses of $\pi^\pm$, $K^\pm$ and $D^\pm$}

\begin{figure}[h!]
\begin{center}
\begin{tabular}{@{}cccc@{}}
\includegraphics*[height=8cm,width=6.5cm,clip=true]{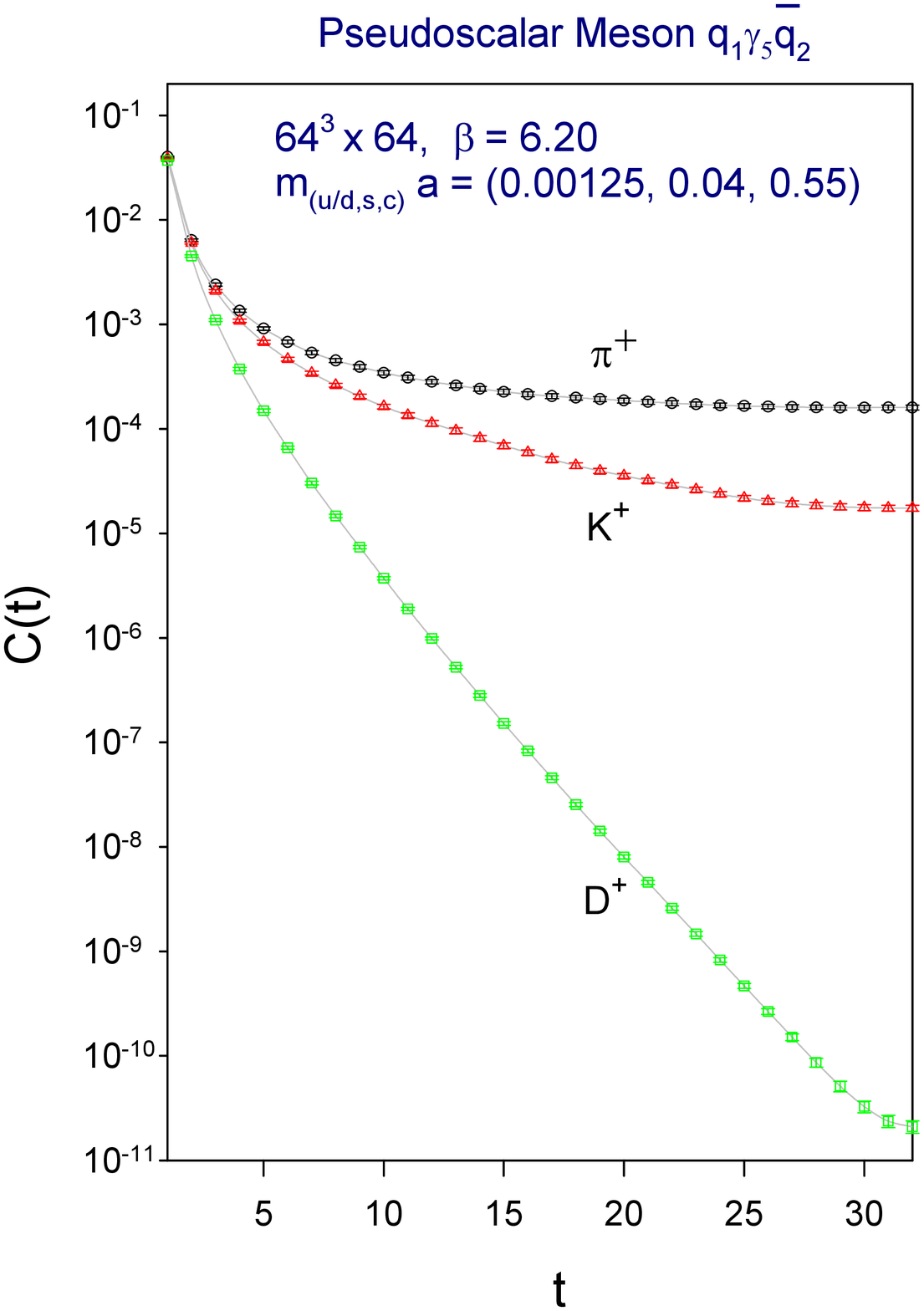}
&
\includegraphics*[height=8cm,width=6.5cm,clip=true]{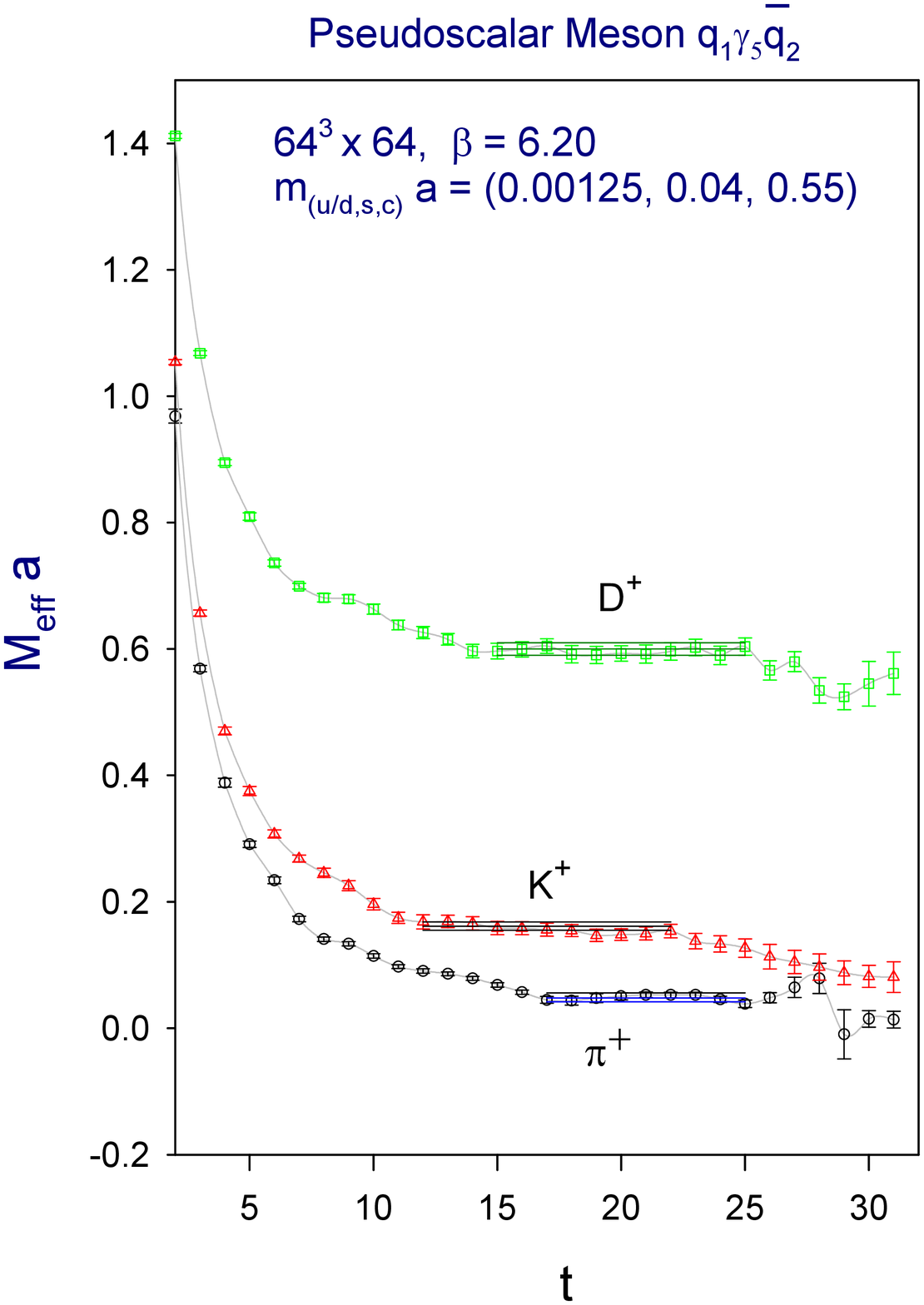}
\\
\end{tabular}
\caption{
The time-correlation function (left panel) and the effective mass (right panel) 
of $ \pi^\pm $, $ K^\pm $, and $ D^\pm $.
}
\label{fig:meson}
\end{center}
\end{figure}

With 50 thermalized HMC trajectories, we sample one configuration every 5 trajectories, 
then we have 10 configurations for the measurement of physical observables.
In Fig. \ref{fig:meson}, the time-correlation functions of $ \ubar \gamma_5 \d $, $ \ubar \gamma_5 \s $, 
and $ \dbar \gamma_5 \c $  are plotted in the left-panel, while their effective massess are plotted 
on the right-panel. 
The fitting results are summarized in Table \ref{tab:meson}. 
Our preliminary result of the masses of $ \pi^\pm $, $ K^\pm $, and $ D^\pm $ are in  
good agreement with the PDG values, suggesting that the quark masses in 
our HMC simulation are close to the physical ones, though in the limit of isospin symmetry.   

\begin{wraptable}{r}{0.58\textwidth}
\caption{The preliminary result of the masses of the lowest-lying pseudoscalar meson states 
         obtained in this work, in comparison with the PDG values.}
\begin{tabular}{ccccc} \\ \toprule
$ \qbar_1 \gamma_5 \q_2 $ & $ [t_1,t_2] $ & $\chi^2$/dof & Mass[MeV] & PDG \\  \midrule
$ \ubar \gamma_5 \d $ & [17,24] & 0.65 & 141(8) & $ \pi^\pm (140) $  \\
$ \ubar \gamma_5 \s $ & [12,22] & 0.81 & 495(7) & $ K^\pm (494) $  \\
$ \dbar \gamma_5 \c $ & [15,22] & 0.63 & 1874(18) & $ D^\pm (1870) $  \\  \bottomrule
\end{tabular}
\label{tab:meson}
\end{wraptable}

\section{Conclusion and Outlook}

This study asserts that it is feasible to simulate lattice QCD with $ (\u, \d, \s, \c) $ 
optimal domain-wall quarks at their physical masses, with good chiral symmetry, and without 
topology freezing. The exact pseudofermion action for one-flavor DWF plays the crucial role 
in the simulation, not only to save the memory consumption such that the entire HMC simulation 
on the $ 64^3 \times 64 \times 16 $ lattice can be fitted into the 128 GB device memory 
of Nvidia DGX-1, but also to enhance the HMC efficiency significantly.

To generate gauge ensembles with physical $ (\u, \d, \s, \c) $ domain-wall quarks, 
we will be in a good position to determine physical quantities, e.g.,  
the hadron mass spectra, the decay constants, and the weak matrix elements,   
as well as to address some subtle nonperturbative physics, 
e.g., the GIM mechanism, and the $ \Delta I = 1/2 $ rule.   
Besides lattice QCD at zero temperature, we are also simulating lattice QCD at finite temperature
with $N_f = 2 + 1 + 1 $ physical domain-wall quarks, for $ T = 130-500 $~MeV,  
which is essential for understanding the role of QCD in the early universe.    

In retrospect, at the beginning of TWQCD domain-wall project in 2008,  
we started with the simulation of 2-flavors QCD on the $ 16^3 \times 32 \times 16 $ lattice 
($ a \sim 0.1 $~fm) with pion mass $ \sim 200 $~MeV \cite{Chiu:2009wh}.  
The simulation can be fitted into one Nvidia GTX-280 card with 1 GB device memory, 
attaining $\sim 150 $~Gflops/s (sustained).
It was hard to imagine that after 10 years, with the advancement in both machine and algorithm, 
now it is feasible to simulate $ N_f = 2 + 1 +1 $ lattice QCD with  
physical domain-wall quarks, and the entire simulation can be fitted into one GPU server on the table top, 
at the speed about one trajectory per day. 
To look forward, the simulation of $ N_f = 1 + 1 + 1 + 1 $ QCD with physical domain-wall quarks 
is already around the corner, while that with $ (\u, \d, \s, \c, \b) $ 
physical domain-wall quarks will be feasible in early 2020s, most likely
requiring more than just one table-top GPU server.

\section*{Acknowledgement}

The author is grateful to Jensen Huang of Nvidia Corporation, for his help to realize this study.   
Also, the author appreciates the technical support from Nvidia Taipei office. 
This work is supported by the Ministry of Science and Technology
(Grant Nos.~107-2119-M-003-008, 105-2112-M-002-016, 102-2112-M-002-019-MY3),
and National Center for Theoretical Sciences (Physics Division).
The author also thanks the members of TWQCD collaboration, in particular, 
Yu-Chih Chen, Han-Yi Chou, Tung-Han Hsieh, and Yao-Yuan Mao,  
for their contributions in the TWQCD code development.

\end{document}